%% file: main.tex
\documentclass[10pt,conference]{IEEEtran}
\IEEEoverridecommandlockouts
\usepackage{amsmath,amssymb,amsfonts}
\usepackage{graphicx}
\usepackage{xcolor}
\usepackage{url}
\usepackage{multirow}
\usepackage{tcolorbox}
\usepackage{booktabs}
\usepackage{tcolorbox}
\usepackage{multirow}
\usepackage{bbding}
\usepackage{utfsym}
\usepackage{ulem}
\usepackage{balance}
\usepackage{wasysym}

\def\eg{\textit{e.g.,} }
\def\ie{\textit{i.e.,} }

\def\method{{\sc SkCoder}\ }

\def\BibTeX{{\rm B\kern-.05em{\sc i\kern-.025em b}\kern-.08em
    T\kern-.1667em\lower.7ex\hbox{E}\kern-.125emX}}

\begin{document}

\title{{\sc SkCoder}: A Sketch-based Approach for Automatic Code Generation}

\author{\IEEEauthorblockN{Jia Li \male}
\IEEEauthorblockA{
Key Lab of High Confidence Software \\
Technology, MoE (Peking University) \\
Beijing, China \\
lijia@stu.pku.edu.cn}
\and
\IEEEauthorblockN{Yongmin Li}
\IEEEauthorblockA{
Key Lab of High Confidence Software \\
Technology, MoE (Peking University) \\
Beijing, China \\
liyongmin@pku.edu.cn}
\and
\IEEEauthorblockN{Ge Li* \thanks{* Corresponding authors}}
\IEEEauthorblockA{
Key Lab of High Confidence Software \\
Technology, MoE (Peking University) \\
Beijing, China \\
lige@pku.edu.cn}
\and
\IEEEauthorblockN{Zhi Jin*}
\IEEEauthorblockA{
Key Lab of High Confidence Software \\
Technology, MoE (Peking University) \\
Beijing, China \\
zhijin@pku.edu.cn}
\and
\IEEEauthorblockN{Yiyang Hao}
\IEEEauthorblockA{
aiXcoder \\
Beijing, China \\
haoyiyang@aixcoder.com}
\and
\IEEEauthorblockN{Xing Hu}
\IEEEauthorblockA{
Zhejiang University \\
Ningbo, China \\
xinghu@zju.edu.cn}
}

\maketitle

\begin{abstract}
Recently, deep learning techniques have shown great success in automatic code generation. 
Inspired by the code reuse, some researchers propose copy-based approaches that can copy the content from similar code snippets to obtain better performance. 
Practically, human developers recognize the content in the similar code that is relevant to their needs, which can be viewed as a \textit{code sketch}. The sketch is further edited to the desired code.
However, existing copy-based approaches ignore the code sketches and tend to repeat the similar code without necessary modifications, which leads to generating wrong results.

In this paper, we propose a sketch-based code generation approach named \method to mimic developers' code reuse behavior. 
Given a natural language requirement, \method retrieves a similar code snippet, extracts relevant parts as a code sketch, and edits the sketch into the desired code.
Our motivations are that the extracted sketch provides a well-formed pattern for telling models ``how to write''. The post-editing further adds requirement-specific details into the sketch and outputs the complete code.
We conduct experiments on two public datasets and a new dataset collected by this work.
We compare our approach to 20 baselines using 5 widely used metrics.
Experimental results show that (1) \method can generate more correct programs, and outperforms the state-of-the-art -- CodeT5-base by 30.30\%, 35.39\%, and 29.62\% on three datasets.
(2) Our approach is effective to multiple code generation models and improves them by up to 120.1\% in Pass@1. (3) We investigate three plausible code sketches and discuss the importance of sketches. (4) We manually evaluate the generated code and prove the superiority of our \method in three aspects.
\end{abstract}


\input{chapter/Introduction}

\input{chapter/motivating_example}

\input{chapter/model}

\input{chapter/study_design}

\input{chapter/result}

\input{chapter/human_evaluation}

\input{chapter/discussion}

\input{chapter/related_work}

\input{chapter/conclusion}

\section*{ACKNOWLEDGMENTS}
This research is supported by the National Key R\&D Program under Grant No. 2021ZD0110303, the National Natural Science Foundation of China under Grant Nos. 62192731, 61751210, 62072007, 62192733, 61832009, and 62192730.

\normalem
\balance
\bibliographystyle{IEEEtran}
\bibliography{reference}

\end{document}

%% file: chapter/Introduction.tex
\section{Introduction}
\label{sec:introduction}

As the complexity and scale of the software continue to grow, developers cost lots of effort to write the source code by hand. Code generation aims to automate this coding process and generate the source code that satisfies a given natural language requirement.
Nowadays, deep learning (DL) techniques have been successfully applied to automatic code generation  \cite{yin2018tranx,sun2020treegen,wang2021codet5}. DL-based models take a natural language (NL) description as the input and output the corresponding source code. The models are trained with a corpus of real NL-code pairs.
During the inference, trained models can automatically generate the desired code for a new NL description. 

Recently, inspired by the code reuse \cite{haefliger2008codereuse}, some researchers \cite{hayati2018redcode,hashimoto2018reedit,parvez2021redcoder} introduce the information retrieval techniques into code generation. They retrieve the similar code and provide it as a supplement to code generation models. The models are trained to copy some content from the similar code and obtain a better performance.
In this paper, we refer to these studies as \textit{copy-based} code generation models.

% Generating the source code from scratch is very challenging. Practically, developers often retrieve a similar code snippet and modify the similar code based on current requirements to obtain the desired code \cite{haefliger2008codereuse}. Some researchers \cite{hayati2018redcode,hashimoto2018reedit,parvez2021redcoder} try to mimic the developers' behavior and propose retrieval-augmented code generation models. They retrieve the similar code and provide it as a supplement to code generation models. Experimental results show that retrieval-augmented code generation models achieve obvious improvements.

% However, previous studies \cite{hayati2018redcode,hashimoto2018reedit,parvez2021redcoder} only take the whole similar code as additional inputs and do not give an insight into the guidance of the similar code.
% In fact, developers will understand the functionality of the similar code, determine reusable contents (\eg API, code structure), and then extract a code \textit{sketch}. 
% The sketch only provides a general code pattern and does not contain requirement-specific details, such as API usage patterns \cite{wang2013APIpattern,niu2017APIpattern}.
% Then, developers add details to the sketch based on current requirements and get a new code snippet.
% But previous retrieval-augmented code generation models ignore the importance of sketches and simply copy the similar code without necessary modifications.
% It greatly limits the performance of code generation models in practical applications.

\begin{figure}[t]
\centering
\includegraphics[width=0.9\linewidth]{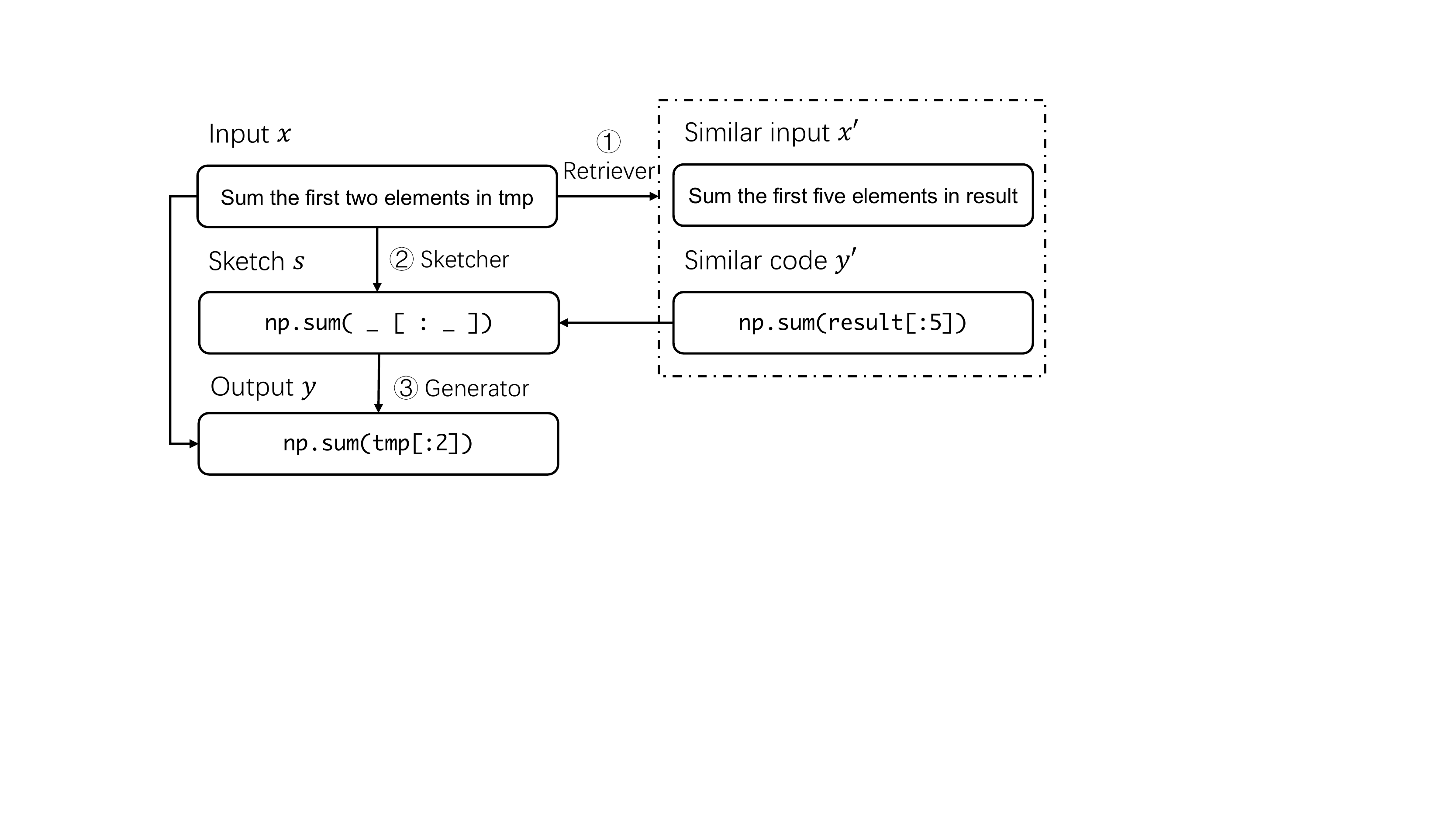}
\caption{The process of reusing the similar code by developers.}
\label{fig:intro_example}
\vspace{-0.5cm}
\end{figure}

\begin{figure*}[t]
\centering
\includegraphics[width=\linewidth]{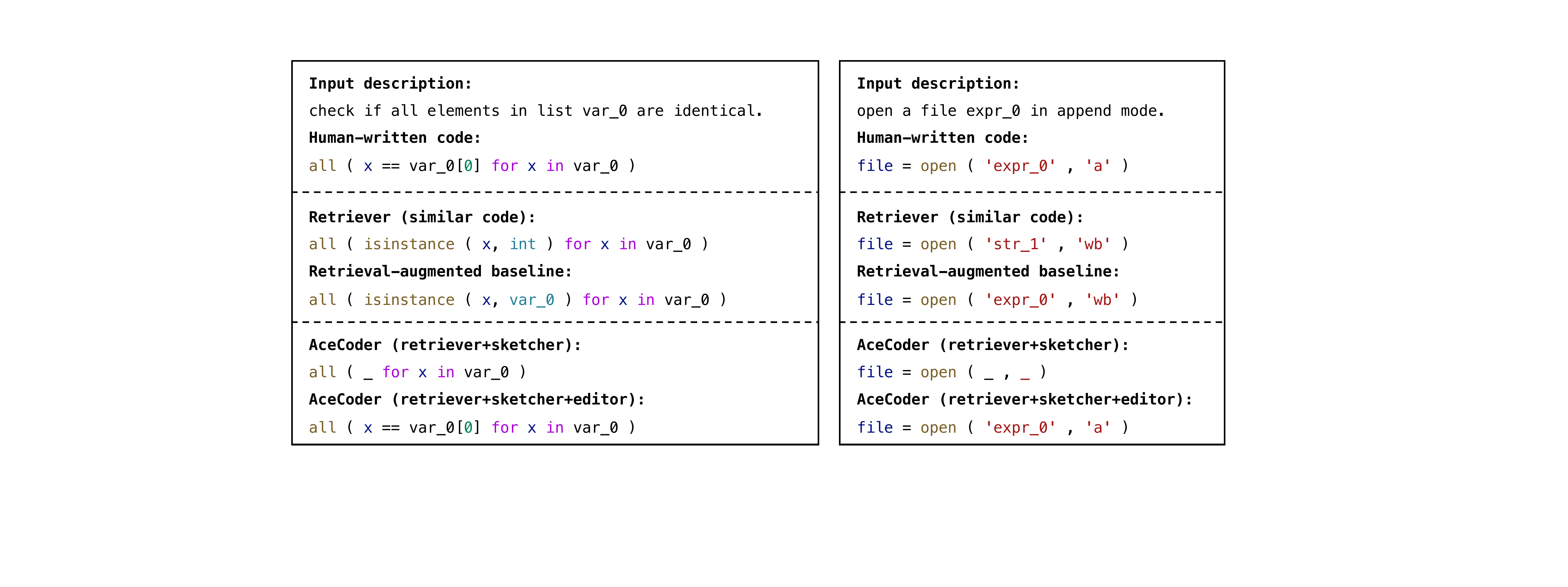}
\caption{The illustration of how developers reuse the similar code. The relevant content in the similar code is highlighted.}
\label{fig:motivating_example}
\end{figure*}

Practically, human developers often make necessary modifications in the similar code instead of simply copying, during the code reuse process \cite{mockus2007codereuse}.
As shown in Figure \ref{fig:intro_example}, developers search for a similar code snippet in open-source communities (\eg Stack Overflow \cite{web:stackoverflow}) and further analyze the relevance of similar code to their requirements.
Then, developers recognize the parts (\ie \texttt{all( \_ for x in \_ )}) that are relevant to their needs and ignore the irrelevant parts (\ie \texttt{x==myList[0]} and \texttt{myList}).
The relevant content can be viewed as a \textit{code sketch}, which specifies a viable code pattern (\eg API usage patterns \cite{wang2013APIpattern,niu2017APIpattern}) to guide developers on how to write their code. 
Next, developers understand the current requirement (\ie check integer) and edit the sketch into the desired code by adding some details (\ie \texttt{isinstance(x,int)}).
In the above pipeline, code sketches play a key role in the code reuse. The sketches denote the knowledge that developers extract from the similar code, and are further reused in the newly-written code.
However, previous copy-based models \cite{hayati2018redcode,parvez2021redcoder} ignore the importance of sketches.
Experimental results show that copy-based models tend to repeat the similar code without necessary modifications and even copy the irrelevant content.

To mimic the above developers' code reuse behavior, we propose a novel \textit{sketch-based} code generation approach, named {\sc SkCoder}.
Different from simply copying in previous copy-based approaches, \method can identify the content in similar code that is relevant to current requirements and further modify those relevant content.
% \method extracts the relevant content from the similar code as a code sketch, and further edits the sketch into the target code based on the NL description.
Our motivations are that code sketches denote the guidance from the similar code that tells models ``how to write'', and NL descriptions express requirements that tell models ``what to write''.
Specifically, \method generates the source code in three steps:
\begin{itemize}
    \item \textbf{Retrieve}. Given an NL description, we use a \textit{retriever} to choose a similar code snippet from a retrieval corpus.
    \item \textbf{Sketch}. Based on the NL description, we use a \textit{sketcher} to extract a code sketch from the similar code.
    \item \textbf{Edit}. We employ an \textit{editor} to edit the sketch based on the NL description and obtain the target code.
\end{itemize}

We conduct extensive experiments to evaluate our {\sc SkCoder}. 
(1) We evaluate \method on two public datasets \cite{ling2016latent}, including HearthStone and Magic.
We employ three widely used evaluation metrics (exact match (EM), BLEU \cite{papineni2002bleu}, and CodeBLEU \cite{ren2020codebleu}). Results demonstrate the impressive performance of our {\sc SkCoder}.
In terms of the EM, \method outperforms state-of-the-art (SOTA) baselines by up to 22.41\% and SOTA copy-based baselines by up to 42.86\%.
(2) We collect a new code generation dataset named AixBench-L that consists of 200k real NL-code pairs. Each test sample is equipped with a set of unit tests. We use Pass@1 and AvgPassRatio to verify the correctness of the generated code.
Results show that \method outperforms SOTA baselines 12.9\% in Pass@1 and 8.49\% in AvgPassRatio.
(3) We conduct an ablation study of our approach on multiple code generation models by gradually adding the retriever and sketcher to these models.
Results prove the contributions of different modules and our \method can substantially improve different models by up to 120.1\% in Pass@1.
(4) We investigate three plausible design choices for code sketches. Results demonstrate the importance of the sketch and our used sketch has a better performance. We also discuss the importance of code sketches through real examples.
(5) We conduct a human evaluation to evaluate the generated code in three aspects, including correctness, code quality, and maintainability. Results show that \method outperforms baselines in all three aspects.

We summarize our contributions in this paper as follows.
\begin{itemize}
    \item To mimic developers' code reuse behavior, we propose a sketch-based code generation approach named {\sc SkCoder}. It extracts a code sketch from the retrieved similar code and further edits the sketch into the target code based on the input description.
    \item We collect a new code generation dataset named AixBench-L that consists of 200k real NL-code pairs. Each test sample is equipped with a set of unit tests to evaluate the correctness of functions.
    \item We conduct extensive experiments on three datasets. Qualitative and quantitative analysis shows the effectiveness of our {\sc SkCoder}. We further investigate different design choices of code sketches and discuss the importance of code sketches.
\end{itemize}

\textbf{Data Availability.}
We open source our replication package \cite{web:SkCoder}, including the datasets and the source code of {\sc SkCoder}, to
facilitate other researchers and practitioners to repeat our work and verify their studies.

%% file: chapter/motivating_example.tex
\section{Motivating examples}
\label{sec:motivating_examples}

In Figure \ref{fig:motivating_example}, we show an example to analyze how developers reuse the similar code and explain our motivations.

\textbf{(1) For an input requirement, the retrieved similar code contains the relevant content and irrelevant parts.}
Given an NL description, developers first retrieve a similar code snippet.
Figure \ref{fig:motivating_example} shows the Top-1 similar code snippet that is retrieved based on the similarity of NL descriptions.
Then, developers understand the implementation details of the similar code and determine which parts are relevant to their requirements.
We can see that the similar code contains lots of relevant content (\ie highlight in Figure \ref{fig:motivating_example}), \eg parameters (\texttt{list1}), control flow statements (\texttt{for x in list1:}), and data flow statements (\texttt{result=result+1}).
Meanwhile, the similar code also contains irrelevant parts, such as the if condition statement (\texttt{if instance(x,int):}).

\begin{figure*}[t]
\centering
\includegraphics[width=\linewidth]{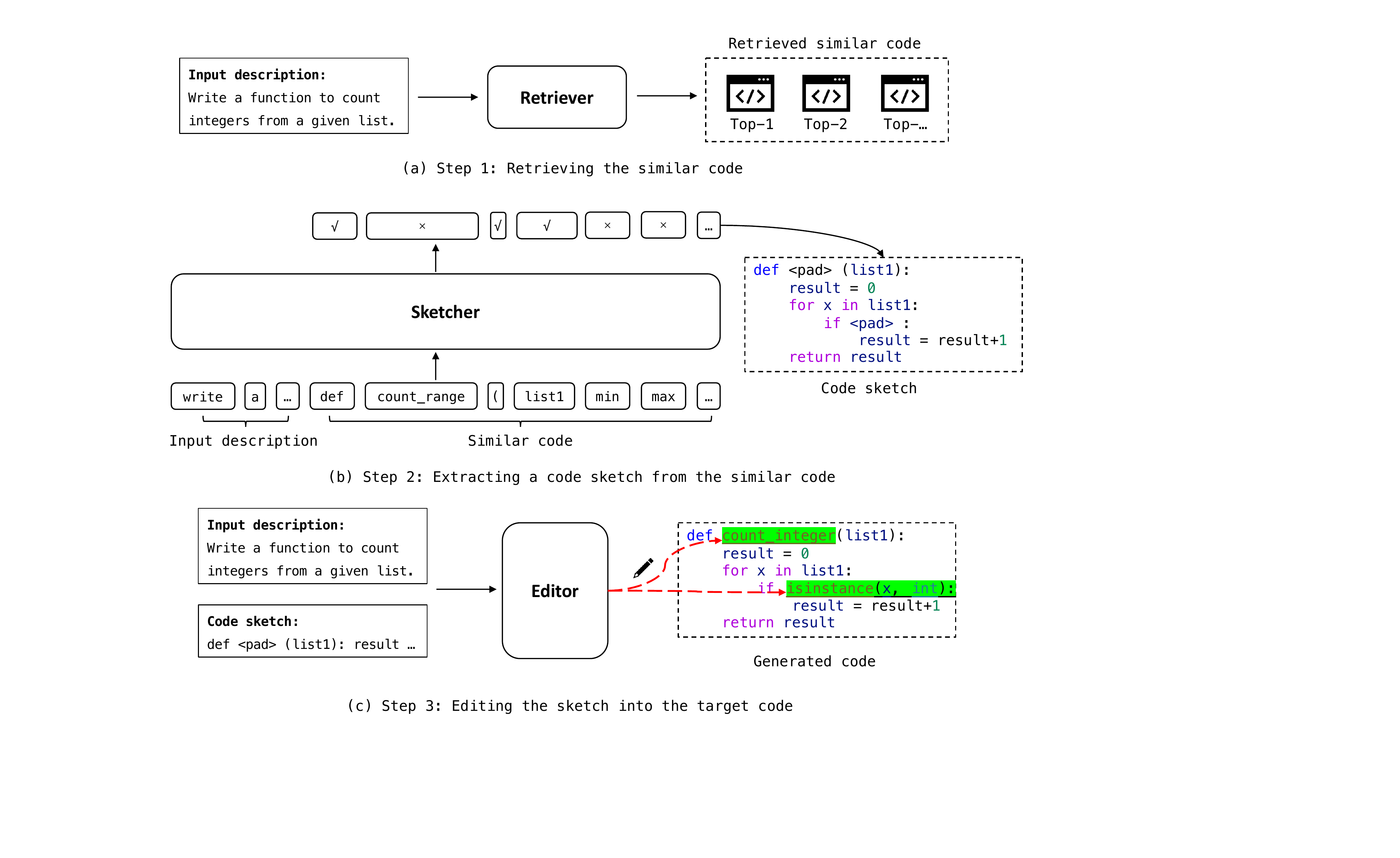}
\caption{The overview of our approach.}
\label{fig:overview}
\end{figure*}

Thus, simply copying from the similar code is inappropriate, which probably causes the generated code contains some irrelevant parts.
We show a wrong output of the SOTA copy-based approach named REDCODER \cite{parvez2021redcoder} in Figure \ref{fig:case_study}.
REDCODER directly copies an incorrect statement from the similar code without necessary modifications.

\textbf{(2) We should extract the relevant content from the similar code as a code sketch.}
Practically, developers will recognize the relevant content from the similar code, ignoring irrelevant parts.
The relevant content can be viewed as a code sketch, which specifies a code pattern to guide developers on how to write the source code.
Figure \ref{fig:motivating_example} shows a sketch extracted from the similar code. The token ``\texttt{\_}'' is a placeholder. 
We can see that the sketch provides a high-level code structure for developers, \ie initializing a counting variable $\rightarrow$ iterating the list and counting $\rightarrow$ returning the counting variable. Some details are replaced by placeholders and elaborated by developers.

Thus, we argue that code sketches are the core of a code reuse process, which denote the valuable knowledge from the similar code and are further reused in the new code.

\textbf{(3) The sketch needs to be edited based on the input description to obtain the target code.}
Code sketches provide code patterns that tell developers ``how to write'', and the NL descriptions express requirements that tell developers ``what to write''.
Thus, developers will edit sketches based on their requirements and obtain the final code.
Figure \ref{fig:motivating_example} shows the final code.
Developers understand requirements (\ie counting elements within a specific range) from the input description and fill in sketches with implementation details, \eg function name (\texttt{count\_range}), if condition statements (\texttt{if min<=x<=max:}).

Based on the above observations, we propose a sketch-based code generation approach to mimic the developers' code reuse behavior. Different from previous copy-based code generation models, our approach contains a sketcher module that can extract the relevant content from the similar code and output a code sketch. Then, we utilize an editor module to edit the sketch into the target code. Through the above pipeline, our approach effectively mines the knowledge from existing high-quality code corpus and transfers the knowledge into newly-written programs.

%% file: chapter/model.tex
\section{Approach}
\label{sec:model}

In this section, we present a sketch-based code generation approach, named {\sc SkCoder}. We formally define the overview of our {\sc SkCoder} and describe the details in the following sections, including three modules and the training details.

\subsection{Overview}
\label{sec:model:overview}

The goal of code generation is to train a model $G(Y|X)$ that predicts a code snippet $Y$ based on an input natural language (NL) description $X$. In this work, we decompose this model into three modules, including a retriever, a sketcher, and an editor. The three modules work in a pipeline as shown in Figure \ref{fig:overview}:
\begin{itemize}
    \item \textbf{Retrieve}. Given an NL description $X$, a retriever selects a similar code snippet $Y'$ from a retrieval corpus.
    \item \textbf{Sketch}. Based on the NL description $X$, a sketcher extracts a code sketch $S$ from the similar code $Y'$.
    \item \textbf{Edit}. An editor edits the sketch $S$ into the target code $Y$ based on the NL description $X$.
\end{itemize}

\subsection{Retriever}
\label{sec:model:retriever}
As shown in Figure \ref{fig:overview} (a), the retriever aims to select similar code snippets from a retrieval corpus based on the input NL description.
Inspired by previous studies \cite{hayati2018redcode,hashimoto2018reedit}, we think that similar code snippets are likely to have similar NL descriptions. Therefore, we take the input description as a query to search for similar descriptions in a retrieval corpus. Then, the corresponding code of similar descriptions is viewed as the similar code.

Specifically, we employ the BM25 score \cite{robertson2009bm25} as our retrieval metric, which is widely used in previous studies \cite{wei2020re2com,li2021editsum,lu2022reacc}. 
BM25 is a bag-of-words retrieval function to estimate the lexical-level similarity of two sentences.
The more similar two sentences are, the higher the value of BM25 scores.
We leverage the open-source search engine Lucene \cite{web:lucene} to build our retriever and use the training set as our retrieval corpus. 

Our motivation is that BM25 and Lucene can bring a nice retrieval accuracy and have low complexity. Considering that the retrieval corpus is often large-scale, a fast retriever is closer to practical applications.
We also notice that there are some more advanced code search approaches \cite{gu2018codesearch,cambronero2019codesearch}, and they can be applied to our approach in a plug-and-play fashion.
Because these approaches have higher complexity, we leave them for further work.

\vspace{-0.2cm}
\subsection{Sketcher}
\label{sec:model:sketcher}

The goal of our sketcher is to extract a code sketch from the similar code based on the input description.
In other words, the sketcher should extract the content that is relevant to the input description and ignore irrelevant parts. 
We consider this procedure as a series of token-level classification actions. 
We first split the similar code into a token sequence. Then, we utilize a neural network to capture relations between the input description and the similar code tokens. For more relevant tokens, the neural network assigns higher weights.
Based on the outputs of the neural network, we further decide whether each token in the similar code is extracted or ignored.
The ignored tokens are replaced by placeholders.
Figure \ref{fig:overview} (b) shows the workflow of our sketcher.

Specifically, we concatenate the NL description $X$ and the similar code $Y'$ into an input sequence and tokenize it. Then, we use a neural encoder $\operatorname{Encoder}(\cdot)$ to convert the input sequence into vector representations $[H;H']$.
\begin{equation}
\begin{aligned}
    X = (x_1, x_2, \dots, x_n) \\
    Y' = (y'_1, y'_2, \dots, y'_m) \\
    [H;H'] = \operatorname{Encoder}([X;Y'])
\end{aligned}
\end{equation}
where $x_i$ and $y'_i$ are the $i$-th token in the NL description and the similar code; $n$ and $m$ are the maximum lengths of the NL description and the similar code.

We further extract vector representations of the similar code and feed them into a linear classification layer. The classification layer will output a probability $p_i$ for each token in the similar code.
If the probability is greater than a threshold $t$, the token is extracted; otherwise, it is replaced with a placeholder (\texttt{<pad>}).
\vspace{-0.1cm}
\begin{equation}
\small
\begin{aligned}
    H' = (h'_1, h'_2, \dots, h'_m) \\
    p_i = \operatorname{softmax}(W_s h'_i + b_s)
\end{aligned}
\end{equation}
\vspace{-0.1cm}
\begin{equation}
\small
s_i = \left\{\begin{array}{ll}
y'_{i} & \text{if} \quad p_{i} > t \\
\texttt{<pad>} & \text {otherwise}
\end{array}\right.
\end{equation}
\vspace{-0.1cm}
\begin{equation}
    S = (s_1, s_2, \dots, s_m)
\end{equation}
where $h'_i$ denotes the vector representation of $i$-th token in the similar code.
$W_s$ and $b_s$ are trainable parameters in the classification layer.
$S$ is the predicted sketch and $s_i$ is the $i$-th token in the sketch. We further merge consecutive placeholders in the sketch into one placeholder.

\vspace{-0.2cm}
\subsection{Editor}
\label{sec:model:editor}

As shown in Figure \ref{fig:overview} (c), our editor treats the sketch as a soft template and generates the target code with the guidance of the input description.
The editor is trained to follow code structures provided by the sketch and add details to some placeholders (\eg \texttt{count\_range}, \texttt{min<=x<=max}). The editor also can generate some necessary components that are not in the sketch, \eg additional parameters (\texttt{min, max}).

In this paper, we employ an encoder-decoder neural network to implement our editor, which has been widely used in code generation \cite{yin2018tranx,sun2020treegen,parvez2021redcoder,wang2021codet5}.
Specifically, we concatenate the NL description and the sketch into an input sequence. The input sequence is transformed into vector representations by an encoder, and a decoder generates the target code based on vector representations.

\vspace{-0.2cm}
\subsection{Training and Testing}
\label{sec:model:training}

Our \method contains three modules: retriever, sketcher, and editor. We employ a deterministic retriever that does not contain trainable parameters. Besides, considering that the sketcher performs non-differentiable hard classifications, the overall approach cannot be trained in an end-to-end fashion.
Thus, we employ a two-stage training strategy (\ie firstly training the sketcher and then training the editor), which is widely used in other fields like code completion \cite{lu2022reacc} and code summarization \cite{li2021editsum}.

\begin{figure}[t]
\centering
\includegraphics[width=0.9\linewidth]{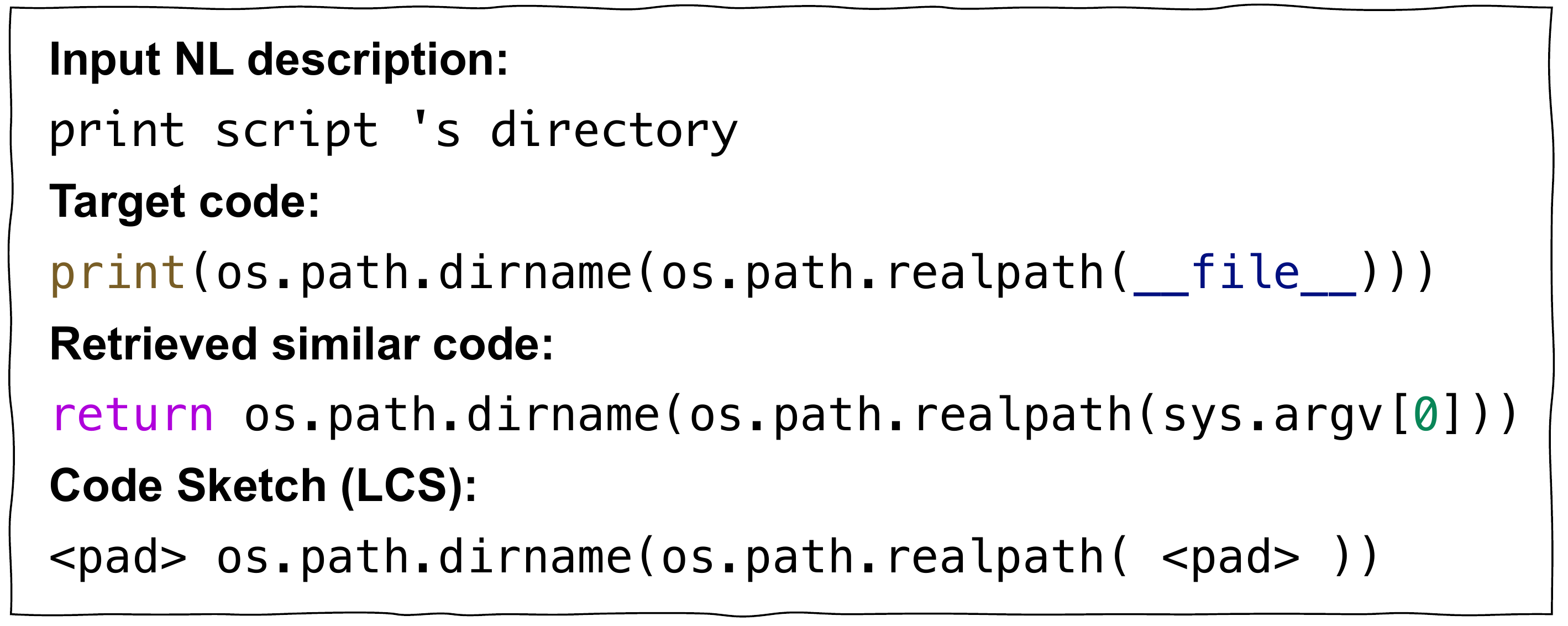}
\vspace{-0.3cm}
\caption{An illustration of our sketch.}
\label{fig:lcs}
\vspace{-0.5cm}
\end{figure}

\subsubsection{Training the sketcher}
The sketcher takes an NL description $X$ and a similar code snippet $Y'$ as inputs and outputs a code sketch $S$.
But existing code generation datasets only contain NL-code pairs $(X,Y)$ without explicit sketches.
Thus, we propose an approach to construct sketches for facilitating the training.
We first pick a dataset and use our retriever to make lots of triples $(X,Y,Y')$.
Then, we treat the longest common subsequence (LCS) \cite{wagner1974lcs} between the similar code $Y'$ and the target code $Y$ as a code sketch $S$.
Figure \ref{fig:lcs} shows an illustration of our sketch. We can see that the LCS effectively keeps reusable parts in the similar code (\eg API and code structures). 
In Section \ref{sec:result}, we experimentally investigate other design choices of sketches and prove the superiority of our used sketch.

Based on the above setting, we can build lots of training triples $(X,Y',S)$. Then, we train our sketcher by minimizing the following loss function:
\begin{equation}
 \mathcal{L}_s = -\sum_{i=1}^m [ \mathbb{I}_i \cdot \log(p_i) + (1-\mathbb{I}_i) \cdot \log(1-p_i)]
\end{equation}
where $p_i$ is a predicted probability that $i$-th token of similar code $y'_i$ is kept in the sketch $S$.
$\mathbb{I}_i$ is an indicator function that outputs 1 when $y'_i$ is in $S$ and outputs 0 when $y'_i$ is not in $S$.

\subsubsection{Training the editor}
The inputs of our editor contain an NL description $X$ and a code sketch $S$, and the output is the target code $Y$. We utilize a retriever to make triples $(X,Y,Y')$ and further use a trained sketcher to predict the code sketches, obtaining lots of training triples $(X,S,Y)$.
We train our editor by minimizing the following loss function:
\begin{equation}
    \mathcal{L}_e = - \sum_{i=1}^m \log P(y_i | X, S, y_{<i})
\end{equation}
where $y_i$ denotes $i$-th token in the target code and $y_{<i}$ is the part of the target code before $y_i$.

\subsubsection{Testing}
After training the sketcher and editor, our \method can be applied to online inference.
Given a new NL description, we use a retriever to search for a similar code snippet from a retrieval corpus. Then, our sketcher extracts a code sketch from the similar code and our editor generates the desired code snippet based on the sketch.

%% file: chapter/study_design.tex
% \vspace{-0.2cm}
\section{Study Design}
\label{sec:study_design}

To assess the effectiveness of our approach, we perform a large-scale study to answer three research questions. In this section, we describe the details of our study, including datasets, metrics, and baselines.

\begin{table}[t]
\centering
\caption{Statistics of the datasets in our experiments.}
\begin{tabular}{lccc}
\toprule
Statistics                 & Hearthstone & Magic & AixBench-L  \\
\midrule
\# Train                   & 533         & 11,969   & 190,000 \\
\# Dev                     & 66          & 664      & 10,000 \\
\# Test                    & 66          & 664      & 175 \\
\midrule
Avg. tokens in description & 27.92       & 59.54  & 27.55 \\
Max. tokens in description & 44          & 174    & 3752 \\
Avg. tokens in code        & 87.14       & 302.44 & 170.74 \\
Max. tokens in code        & 407         & 2395  & 25237 \\
\bottomrule
\end{tabular}
\label{tab:dataset}
\vspace{-0.4cm}
\end{table}

\subsection{Research Questions}
\label{sec:study_design:RQ}

Our study aims to answer three research questions (RQ). In RQ1, we compare our \method to SOTA code generation models on three representative datasets. In RQ2, we conduct an ablation study to prove the contributions of different modules. In RQ3, we investigate different design choices of code sketches and validate the effectiveness of our design.

\textbf{RQ1: How does \method perform compared to SOTA baselines?}
We train our \method with three representative datasets. Then, we use multiple metrics to evaluate the \method and compare it to existing SOTA code generation baselines.

\textbf{RQ2: What are the contributions of different modules in our approach?}
Our \method consists of three modules: a retriever, a sketcher, and an editor. We assess the contributions of different modules by gradually adding them to a base model. We select multiple neural networks as the base models and aim to verify that our approach is effective to different network architectures.

\textbf{RQ3: What is the better design choice of the sketcher?}
In Section \ref{sec:model:training}, we treat the longest common subsequence (LCS) as the code sketch. In this RQ, we provide other design choices of the sketch and compare them to our design.

\vspace{-0.2cm}
\subsection{Datasets}
\label{sec:study_design:dataset}
We conduct experiments on two public datasets (\ie HearthStone in Python and Magic in Java) collected by Ling et al. \cite{ling2016latent} and a new Java dataset named AixBench-L collected by this work.

\textbf{HearthStone and Magic datasets} are proposed for the automatic code generation for cards in games. Each sample is composed of a semi-structural description and a human-written program. The description comes with several attributes such as card name, and card type, as well as a natural language description for the effect of the card.
We follow previous work \cite{yin2018tranx,sun2020treegen} to pre-process the two datasets, and the statistic is listed in Table \ref{tab:dataset}. 

\textbf{AixBench-L} is a function-level code generation benchmark and is an augmented version of the public AixBench benchmark \cite{hao2022aixbench}. 
We treat the original AixBench as the test data and collect lots of NL-code pairs from Github \cite{web:github} as the train and dev data. 
Specifically, we mined Java open-source projects with at least 30 stars from GitHub, and avoid projects containing test data. From mined projects, we remove auto-generated functions and extract functions (i) having an English docstring; (ii) having $<$1024 tokens and $>$1 lines. Finally, we select 200k samples from mined projects and randomly split them into train data and valid data. We consider all mined projects as the retrieval corpus.
The statistic is shown in Table \ref{tab:dataset}. Each test sample contains a functionally independent and well-described natural language description, a signature of the target function, and a set of unit tests that verify the correctness of the function. Following previous work \cite{hao2022aixbench}, we take the natural language description and the function signature as models' inputs.

\vspace{-0.2cm}
\subsection{Metrics}
\label{sec:study_design:metric}

On HearthStone and Magic datasets, we view human-written programs as the ground-truth, and employ three widely used metrics to evaluate the similarity of the generated code and the ground-truth \cite{yin2018tranx,sun2020treegen,wang2021codet5}.
\begin{itemize}
    \item \textbf{Exact match (EM)} is the percentage of the generated code that has the same token sequence as the ground-truth.
    \item The \textbf{BLEU} score \cite{papineni2002bleu} is used to measure the token-level similarity between the generated code and the ground-truth. Specifically, it calculates the $n$-gram similarity and can be computed as:
    \begin{equation}
    \small
    \text{BLEU} = \text{BP} \cdot \exp \left(\sum_{n=1}^{N} w_{n} \log p_{n}\right)
     \end{equation}
    where $p_n$ is the $n$-gram matching precision scores, $N$ is set to 4 in our experiments. $\text{BP}$ is a brevity penalty to prevent very short generated code.
    \item The \textbf{CodeBLEU} score \cite{ren2020codebleu} is a variant of the BLEU score. It specializes in the source code and considers syntactic and semantic matches based on the code structure in addition to the n-gram match.
\end{itemize}

The test data in AixBench-L does not contain human-written programs. We have to omit metrics (e.g., EM, BLEU) requiring ground-truths. Following previous work \cite{hao2022aixbench}, we use unit tests to evaluate the correctness of generated programs. Specifically, we employ the following metrics:
\begin{itemize}
    \item \textbf{Pass@1} is the percentage of the generated code that passes all unit tests. It has been widely used in previous studies \cite{zan2022cert,web:codeparrot,web:gpt-cc}.
    \item \textbf{AvgPassRatio} denotes the average test cases pass ratio and can be calculated like this:
    \begin{equation}
    \begin{aligned}
    \text {AvgPassRatio} &=\frac{1}{T} \sum_i^T \text {PassRatio}_i \\
    \text {PassRatio}_i &=\frac{\text{Count}_{i, \text{pass}}}{\text{Count}_{i, \text{total}}}
    \end{aligned}
    \end{equation}
    where ${\text{Count}}_{i,\text{pass}}$ and ${\text{Count}}_{i,\text{total}}$ are the number of passed test cases and the total number of test cases in $i$-th test sample, respectively. $T$ is the size of test data.
\end{itemize}

\subsection{Baselines}
\label{sec:study_design:baseline}

We select 20 recently proposed code generation models as baselines. They can be divided into three categories: sequence-based baselines, tree-based baselines, and pre-trained baselines.

The sequence-based baselines treat the source code as plain text and directly generate a code token sequence:
\begin{itemize}
    \item \textbf{RNN} \cite{zaremba2014rnn} is a classic neural network in source code processing. We utilize the RNN to implement a vanilla encoder-decoder code generation model as the baseline.
    \item \textbf{Transformer} \cite{vaswani2017transformer} is a popular encoder-decoder model and has obtained promising results in code generation and code completion tasks \cite{lu2022reacc}.
    \item \textbf{LPN} \cite{ling2016latent} and \textbf{ReEdit} \cite{hashimoto2018reedit} are RNN-based code generation models. LPN proposes a structured attention mechanism to handle the semi-structural inputs. ReEdit introduces a retrieved similar program as an additional input.
\end{itemize}

The tree-based baselines directly generate a parsed tree (\eg abstract syntax tree) of the source code.
The generated tree is further converted to the source code.
\begin{itemize}
    \item \textbf{Seq2Tree} \cite{dong2016seq2tree} is a pioneer tree-based work that proposes a attention-enhanced code generation model.
    \item \textbf{TRANX} \cite{yin2018tranx} is a representative tree-based code generation model that can map an NL description into a tree using a series of tree construction actions.
    \item \textbf{ASN} \cite{rabinovich2017asn} utilizes a dynamically-determined decoder to efficiently generate a tree.
    \item \textbf{TreeGen} \cite{sun2020treegen} incorporates grammar rules and tree structures into the Transformer. It significantly outperforms previous RNN-based code generation models.
    \item \textbf{ReCode} \cite{hayati2018redcode} is a variant of the TRANX, which can copy $n$-gram actions from the tree of a similar program.
\end{itemize}

The pre-trained baselines are first pre-trained with a large-scale code corpus and then fine-tuned with code generation datasets.
Nowadays, pre-trained code generation models have achieved SOTA results on many code generation datasets.
\begin{itemize}
    \item \textbf{CodeBERT} \cite{feng2020codebert} and \textbf{GraphCodeBERT} \cite{guo2020graphcodebert} are two encoder-only pre-trained models. They mainly apply the pre-training techniques for natural languages to the source code. We add a six-layer transformer decoder along with the two models, to support code generation. Both models contain 175 million parameters.
    \item \textbf{CodeGPT} \cite{lu2021codexglue} and \textbf{CodeParrot} \cite{web:codeparrot} are two decoder-only pre-trained models. They are derived from the GPT-2 \cite{radford2019gpt-2} and are continually pre-trained with the code. Both models contain 124 million parameters.
    \item \textbf{PyCodeGPT} \cite{zan2022cert} and \textbf{GPT-CC} \cite{web:gpt-cc} are two decoder-only pre-trained models. They are initialized with the GPT-Neo \cite{black10gpt-neo} and are continually pre-trained with a large-scale code corpus in Python.
    Both models contain 110 million parameters.
    \item \textbf{CERT-PyCodeGPT} \cite{zan2022cert} is a variant of the PyCodeGPT. It first predicts a sketch based on the NL description and further generates the complete code based on the sketch. We follow instructions in the original paper and train a CERT-PyCodeGPT (220M) in our experimental datasets.
    \item \textbf{CodeGen} \cite{nijkamp2022codegen} is a decoder-only pre-trained model. It casts code generation as a multi-turn conversation between a user and a system. In this paper, we use the CodeGen-Mono-350M version.
    \item \textbf{REDCODER} \cite{parvez2021redcoder} is a encoder-decoder pre-trained model. It provides multiple similar code snippets as a supplement to a pre-trained code generation model. We use GraphCodeBERT to initialize the retriever and employ PLBART-base \cite{ahmad2021plbart} to initialize the generator. The full REDCODER contains 315 million parameters.
    \item \textbf{CodeT5-small} and \textbf{CodeT5-base} \cite{wang2021codet5} are two encoder-decoder pre-trained models. They propose an identifier-aware pre-training task and have achieved SOTA results on many code generation datasets. CodeT5-small contains 60 million parameters and CodeT5-base consists of 220 million parameters.
    % \item \textbf{Github Copliot} \cite{web:copliot} is a powerful decoder-only pre-trained model. It can suggest individual lines and whole functions based on input context. It is available as an extension for many IDEs (\eg VSCode \cite{web:vscode}).
\end{itemize}

\vspace{-0.2cm}
\subsection{Implementation Details}
\label{sec:study_design:implementation}
The implementation details of our \method are as follows:

\begin{itemize}
    \item \textbf{Retriever.} We use the open-source search engine - Lucene \cite{web:lucene} to build the retriever. The retrieval metric is the BM25 score. For HearthStone and Magic, the retrieval corpus is its training data. Note that we exclude the ground truths from the outputs of our retriever.
    \item \textbf{Sketcher.} We implement the sketcher with a 12-layer Transformer encoder. Its network architecture follows previous studies \cite{feng2020codebert,guo2020graphcodebert}. We initialize the sketcher using pre-trained weights of GraphCodeBERT \cite{guo2020graphcodebert}.
    \item \textbf{Editor.} The editor is an encoder-decoder Transformer, and the encoder and decoder both contain 12 Transformer layers.
    The editor follows the network architecture in the work \cite{wang2021codet5} and is initialized with pre-trained weights of CodeT5-base \cite{wang2021codet5}.
    \item \textbf{Training \& Testing.} We train the \method with two NVIDIA A100 GPUs. The batch size is set to 32. During training, we use Top-5 similar code snippets to build the training data of our sketcher and editor. In the inference, we only use the Top-1 similar code, 
    employ the beam search, and set the beam size to 10.
\end{itemize}

Note that initializing using pre-trained weights is common in previous studies \cite{guo2020graphcodebert,parvez2021redcoder,zan2022cert,web:codeparrot,web:gpt-cc} and can effectively improve the performance of models.
To make a fair comparison, we also reuse the pre-trained weights in our experiments. 

%% file: chapter/result.tex
\vspace{-0.2cm}
\section{Results and Analyses}
\label{sec:result}

In our first research question, we evaluate the performance of our \method with respect to previous code generation approaches.

% \begin{tcolorbox}[size=title]
% \textbf{RQ1: How does \method perform compared to SOTA baselines?}
% \end{tcolorbox}

\noindent \textbf{RQ1: How does \method perform compared to SOTA baselines?}

\begin{table}[t]
\caption{Results on the HearthStone dataset (Python). ``*'' represents the copy-based baselines.}
\centering
\resizebox{\linewidth}{!}{
\begin{tabular}{llccc}
\toprule
\textbf{Type}       & \textbf{Approach}    & \textbf{EM} & \textbf{BLEU}  & \textbf{CodeBLEU} \\
\midrule
                                & Retriever module & 0  & 57.56 & 56.58    \\
\midrule
\multirow{4}{*}{Sequence-based} & LPN              & 6.10   & 67.10  & --         \\
                                & RNN              & 3.03   & 64.53  & 58.56         \\
                                & Transformer      & 3.03   & 62.46 & 51.63         \\
                                & ReEdit *         & 9.10  & 70.00  & --  \\
\midrule
\multirow{5}{*}{Tree-based}     & Seq2Tree         & 1.50 & 53.40  & --          \\
                                & TRANX            & 16.20  & 75.80  & --          \\
                                & ASN              & 18.20  & 77.60   & --         \\
                                & ReCode *         & 19.60  & 78.40   & --          \\
                                & TreeGen          & 25.80   & 79.30   & --         \\
\midrule
\multirow{11}{*}{Pre-trained}   & CodeBERT         & 3.03   & 66.50  & 59.39         \\
                                & GraphCodeBERT    & 3.03   & 66.32  & 58.87         \\
                                & CodeGPT          & 15.15   & 80.90 & 66.69         \\
                                & GPT-CC        & 15.15  & 74.58  & 63.95         \\
                                & CodeParrot       & 19.70   & 76.99  & 65.40         \\
                                & PyCodeGPT         & 24.24  & 81.03  & 68.70         \\
                                & CERT-PyCodeGPT   & 16.67    & 78.91   & 67.73          \\
                                & CodeGen          & 24.24   & 78.80  & 67.43         \\
                                & REDCODER *       & 21.21   & 80.08    & 67.31          \\
                                & CodeT5-small     & 21.20   & 77.91 & 64.60         \\
                                & CodeT5-base      & 25.84   & 81.28  & 68.42          \\
                                & \method        & \textbf{30.30 ( $\uparrow$ 17.26\%)}    & \textbf{83.12 ( $\uparrow$ 2.26\%)}      & \textbf{70.97 ( $\uparrow$ 3.73\%)}         \\
\bottomrule
\end{tabular}}
\label{tab:result_hs}
\vspace{-0.3cm}
\end{table}

\begin{table}[t]
\centering
\caption{Results on the Magic dataset (Java). We omit some baselines as they cannot be applied to the Java language.}
\resizebox{\linewidth}{!}{
\begin{tabular}{llccc}
\toprule
\textbf{Type}       & \textbf{Approach}    & \textbf{EM} & \textbf{BLEU}  & \textbf{CodeBLEU} \\
\midrule
                                & Retriever module & 0  & 53.64 & 64.23    \\
\midrule
\multirow{3}{*}{Sequence-based} & LPN              & 4.80   & 61.40  & --         \\
                                & RNN              & 16.26   & 71.96  & 61.83         \\
                                & Transformer      & 12.20   & 73.10 & 66.61         \\
\midrule
\multirow{8}{*}{Pre-trained}   & CodeBERT         & 19.42  & 78.69  & 71.73         \\
                                & GraphCodeBERT    & 27.41   & 82.33  & 74.76         \\
                                & CodeGPT          & 27.40   & 78.68 & 70.04         \\
                                & REDCODER *       & 9.79   & 58.81    & 50.38          \\
                                & CodeT5-small     & 26.95   & 78.38 & 71.11         \\
                                & CodeT5-base      & 28.91   & 80.46  & 73.11          \\
                                & \method        & \textbf{35.39 ($\uparrow$ 22.41\%)}    & \textbf{85.39 ($\uparrow$ 6.13\%)}      & \textbf{82.42 ($\uparrow$ 10.27\%)}         \\
\bottomrule
\end{tabular}}
\label{tab:result_magic}
\end{table}

\begin{table}[t]
\centering
\caption{Results on the AixBench-L dataset (Java). We omit some baselines as they cannot be applied to the Java language.}
\resizebox{\linewidth}{!}{
\begin{tabular}{llcc}
\toprule
Type       & Approach    & Pass@1 & AvgPassRatio   \\
\midrule
                                & Retriever module & 2.86  & 7.93     \\
\midrule
\multirow{2}{*}{Sequence-based} & RNN              & 4.00   & 13.33           \\
                                & Transformer      & 6.29   &  12.43        \\
\midrule
\multirow{7}{*}{Pre-trained}   & CodeBERT          & 9.14  & 23.35           \\
                                & GraphCodeBERT    & 10.86   & 24.99           \\
                                & CodeGPT          & 17.71   & 35.67          \\
                                & REDCODER *       & 16.00   &  33.14             \\
                                & CodeT5-small     & 12.57   & 25.11          \\
                                & CodeT5-base      & 15.43   & 24.53            \\
                                & \method        & \textbf{20.00 ($\uparrow$ 12.9\%)}    & \textbf{38.70 ($\uparrow$ 8.49\%)}        \\
\bottomrule
\end{tabular}}
\label{tab:result_AixBench}
\vspace{-0.5cm}
\end{table}

\begin{table*}[t]
\centering
\caption{The results of ablation study.}
\resizebox{\linewidth}{!}{
\begin{tabular}{ccccccccccc}
\toprule
\multirow{2}{*}{Editor} & \multirow{2}{*}{Retriever} & \multirow{2}{*}{Sketcher} & \multicolumn{3}{c}{HearthStone}                                                  & \multicolumn{3}{c}{Magic}   & \multicolumn{2}{c}{AixBench-L}  \\
                        &                            &                           & \multicolumn{1}{c}{EM} & \multicolumn{1}{c}{BLEU} & \multicolumn{1}{c}{CodeBLEU} & \multicolumn{1}{c}{EM} & \multicolumn{1}{c}{BLEU} & \multicolumn{1}{c}{CodeBLEU} & \multicolumn{1}{c}{Pass@1} & \multicolumn{1}{c}{AvgPassRate} \\ \midrule
\multirow{3}{*}{RNN}    & \usym{2715}   & \usym{2715}    & 3.03   & 64.53   & 57.56     & 16.26    & 71.96   & 61.83 & 4.00 & 13.33    \\
                        & \usym{1F5F8}   & \usym{2715}   & 3.03 ($\uparrow$ 0\%)   & 68.39   & 59.12     & 16.51 ($\uparrow$ 1.54\%)   & 72.79   & 63.82  & 5.14 ($\uparrow$ 28.5\%) & 10.61   \\
                        & \usym{1F5F8}   & \usym{1F5F8}    & \textbf{4.54 ($\uparrow$ 49.83\%)}   & \textbf{71.50}   & \textbf{61.76}     & \textbf{17.91 ($\uparrow$ 10.15\%)}    & \textbf{73.72}   & \textbf{65.04} & \textbf{8.57 ($\uparrow$ 114.3\%)} & \textbf{13.42 ($\uparrow$ 2.7\%)}    \\ \midrule
\multirow{3}{*}{CodeT5-small}    & \usym{2715}   & \usym{2715}    & 21.20   & 77.91   & 64.60     & 26.95    & 78.38   & 71.11  & 12.57 & 25.11   \\
                        & \usym{1F5F8}   & \usym{2715}    & 27.86 ($\uparrow$ 31.42\%)   & 79.84   & 68.76     & 31.73 ($\uparrow$ 17.74\%)    & 80.85   & 77.10  & 14.29 ($\uparrow$ 13.68\%) & 26.06 ($\uparrow$ 3.78\%)   \\
                        & \usym{1F5F8}   & \usym{1F5F8}    & \textbf{30.30 ($\uparrow$ 42.90\%)}   & \textbf{83.08}   & \textbf{69.35}     & \textbf{33.89 ($\uparrow$ 25.75\%)}    & \textbf{85.15}   & \textbf{80.08}    & \textbf{18.29 ($\uparrow$ 45.51\%)} & \textbf{34.05 ($\uparrow$ 35.6\%)} \\ \midrule
\multirow{3}{*}{CodeT5-base}    & \usym{2715}   & \usym{2715}    & 25.24   & 81.28   & 68.42     & 28.91    & 80.46   & 73.11  & 15.43 & 24.52   \\
                        & \usym{1F5F8}   & \usym{2715}    & 27.81 ($\uparrow$ 10.18\%)   & 82.06   & 69.35     & 32.43 ($\uparrow$ 12.18\%)    & 83.11   & 78.97  & 17.71  ($\uparrow$ 14.78\%) & 34.75 ($\uparrow$ 41.72\%)   \\
                        & \usym{1F5F8}   & \usym{1F5F8}    & \textbf{30.30 ($\uparrow$ 20.05\%)}   & \textbf{83.12}   & \textbf{70.97}     & \textbf{35.39 ($\uparrow$ 22.41\%)}    & \textbf{85.39}   & \textbf{80.62}  & \textbf{20.00 ($\uparrow$ 29.62\%)} & \textbf{38.70 ($\uparrow$ 57.83\%)}    \\
\bottomrule
\end{tabular}}
\label{tab:ablation_study}
\vspace{-0.3cm}
\end{table*}

\textbf{Setup.} We evaluate baselines (Section \ref{sec:study_design:baseline}) and our \method on three code generation datasets  (Section \ref{sec:study_design:dataset}). 
The evaluation metrics are described in Section \ref{sec:study_design:metric}, \ie the EM, BLEU, CodeBLEU, Pass@1, and AvgPassRatio.
For all metrics, higher scores represent better performance.

\textbf{Results.} Table \ref{tab:result_hs}, Table \ref{tab:result_magic} and Table \ref{tab:result_AixBench} show the experimental results on three datasets, respectively. ``--'' denotes that the models have not been evaluated using this metric, to the best of our knowledge. 
``*'' represents the copy-based baselines, which also use the retrieved similar code. The percentages in parentheses are the relative improvements compared to the strongest baselines.
On Magic and AixBench-L datasets, we omit some baselines because they are designed for specific languages and cannot work in the Java dataset. 

\textbf{Analyses.} 
\uline{(1) Our \method achieves the best results among all baselines.} Our \method can generate more correct programs.
Compared to the SOTA model - CodeT5-base, \method outperforms it by up to 22.41\% in EM and 29.62\% in Pass@1. Note that the EM and Pass@1 are very strict metrics and are hard to be improved. The significant improvements prove the superiority of our \method in automatic code generation. 
\uline{(2) The retrieved code is beneficial to code generation.}
Our retriever module performs well in the BLEU and CodeBLEU, but it is poor in the EM and Pass@1.
It validates our motivation that the similar code contains lots of reusable contents and irrelevant parts.
By introducing the retrieved code, code generation models can be further improved. For example, on the HearthStone dataset, ReEdit improves its base model (\ie RNN) by up to 200\%, and ReCode improves its base model (\ie TRANX) by up to 20.99\%.
\uline{(3) Our \method outperforms the SOTA copy-based baselines.}
The SOTA copy-based baseline is the REDCODER, which uses multiple similar code snippets to augment code generation models. While our \method only uses the Top-1 similar code.
Compared to the REDCODER, \method improves it by 42.86\% in EM, 25\% in Pass@1, and 16.78\% in AvPassRatio.
This is because REDCODER is likely to repeat the similar code without necessary modifications. 
While our \method utilizes a sketcher to extract the relevant content as a sketch, ignoring irrelevant parts. The sketch is further edited based on the input description.
Thus, our \method is closer to developers' code reuse behavior and can generate more correct programs.

On the HearthStone and Magic datasets, we notice that the improvements on the EM are higher than those on other metrics.
We carefully compare the output of different models and find that baselines and our \method all can correctly generate the body of programs. But baselines often err on some details, such as parameters.
Thus, our \method can generate more exactly correct programs, and achieve lower improvements on n-gram similarity metrics (\ie BLEU and CodeBLEU).
The results also verify that compared to generating the code from scratch, editing a well-formed sketch is easier to generate the correct code.

\begin{tcolorbox}[size=title]
\textbf{Answer to RQ1:} \method achieves the best results among all baselines. In particular, \method generates 30.30\%, 35.39\%, and 20\% correct programs on three datasets, outperforming the SOTA code generation models by 17.26\%, 22.41\%, and 12.9\%. The significant improvements prove our sketch-based approach is more promising in automatic code generation.
\end{tcolorbox}

In RQ2, we aim to figure out the contributions of different modules in our {\sc SkCoder}. Besides, we plan to investigate the effectiveness of our approach on different code generation models.

\noindent \textbf{RQ2: What are the contributions of different modules in our approach?}

\textbf{Setup.} In this RQ, we select three code generation models as the base editor, including RNN, CodeT5-small, and CodeT5-base. They cover mainstream network architectures, \ie RNN, Transformer, and pre-trained models.
For each editor, we conduct an ablation study by gradually adding the retriever and sketcher.

\textbf{Results.} The experimental results is shown in Table \ref{tab:ablation_study}. 
$\usym{1F5F8}$ and $\usym{2715}$ represent adding and removing corresponding modules, respectively.
An individual editor is just a vanilla code generation model that maps an NL description to the source code.
After adding a retriever, the model takes the retrieved code as an additional input.
After further introducing a sketcher, the model is our sketch-based approach.

\textbf{Analyses.}
(1) \uline{All three modules are necessary to perform the best.}
After adding a retriever, the performance of all models is improved. For example, on HearthStone, the retriever brings a 10.18\% improvement in the EM for the CodeT5-base. It validates that the retrieved code contains lots of valuable information that benefits code generation models.
After introducing a sketcher, all models obtain better results. For example, on the HearthStone, the CodeT5-base is improved by 20.05\% in the EM.
It proves that compared to copying from the retrieved code, our sketch-based code generation approach can better mine the knowledge in the retrieved code.
(2) \uline{Our approach is effective to multiple code generation models.}
As shown in Table \ref{tab:ablation_study}, our approach supports different code generation models and brings obvious improvements.
Specifically, in terms of the Pass@1, our approach improves the RNN by up to 114.3\%, the CodeT5-small by 45.31\%, and the CodeT5-base by 29.62\%.
In the future, our approach can be used to enhance more powerful code generation models.

\begin{tcolorbox}[size=title]
\textbf{Answer to RQ2:} All three modules are essential for the performance of our approach. Besides, our approach is effective to different code generation models and improves them by 114.3\%, 45.31\%, and 29.62\% in Pass@1. 
\end{tcolorbox}

\begin{table}[t]
\caption{The performance of different sketchers.}
\vspace{-0.2cm}
\resizebox{\linewidth}{!}{
\begin{tabular}{lcccccc}
\toprule
\multirow{2}{*}{Approach} & \multicolumn{3}{c}{HearthStone}                                                  & \multicolumn{3}{c}{Magic}                                                        \\
                          & \multicolumn{1}{c}{EM} & \multicolumn{1}{c}{BLEU} & \multicolumn{1}{c}{CodeBLEU} & \multicolumn{1}{c}{EM} & \multicolumn{1}{c}{BLEU} & \multicolumn{1}{c}{CodeBLEU} \\ \midrule
Without sketcher          & 27.81    & 82.06   & 69.35   & 32.43   & 82.01   & 78.87      \\
Sketcher-1                & 27.93 ($\uparrow$ 0.43\%)    & 82.39   & 70.81   & 33.06 ($\uparrow$ 1.94\%)  & 83.04   & 80.15     \\
Sketcher-2                & 29.03 ($\uparrow$ 4.39\%)    & 82.77   & 70.27   & 34.46 ($\uparrow$ 5.95\%)   & 83.91   & 80.19      \\
Our Sketcher              & \textbf{30.30 ($\uparrow$ 9.13\%)}    & \textbf{83.12}   & \textbf{70.97}   & \textbf{35.39 ($\uparrow$ 9.13\%)}   & \textbf{85.39}   & \textbf{80.62}      \\
\bottomrule
\end{tabular}}
\label{tab:sketch}
\vspace{-0.3cm}
\end{table}

Code sketches are not explicitly defined in existing datasets and how to build a sketch is an open question, 
Thus, we design several plausible design choices for the sketcher and investigate which one is better.

\begin{figure}[t]
\centering
\includegraphics[width=0.8\linewidth]{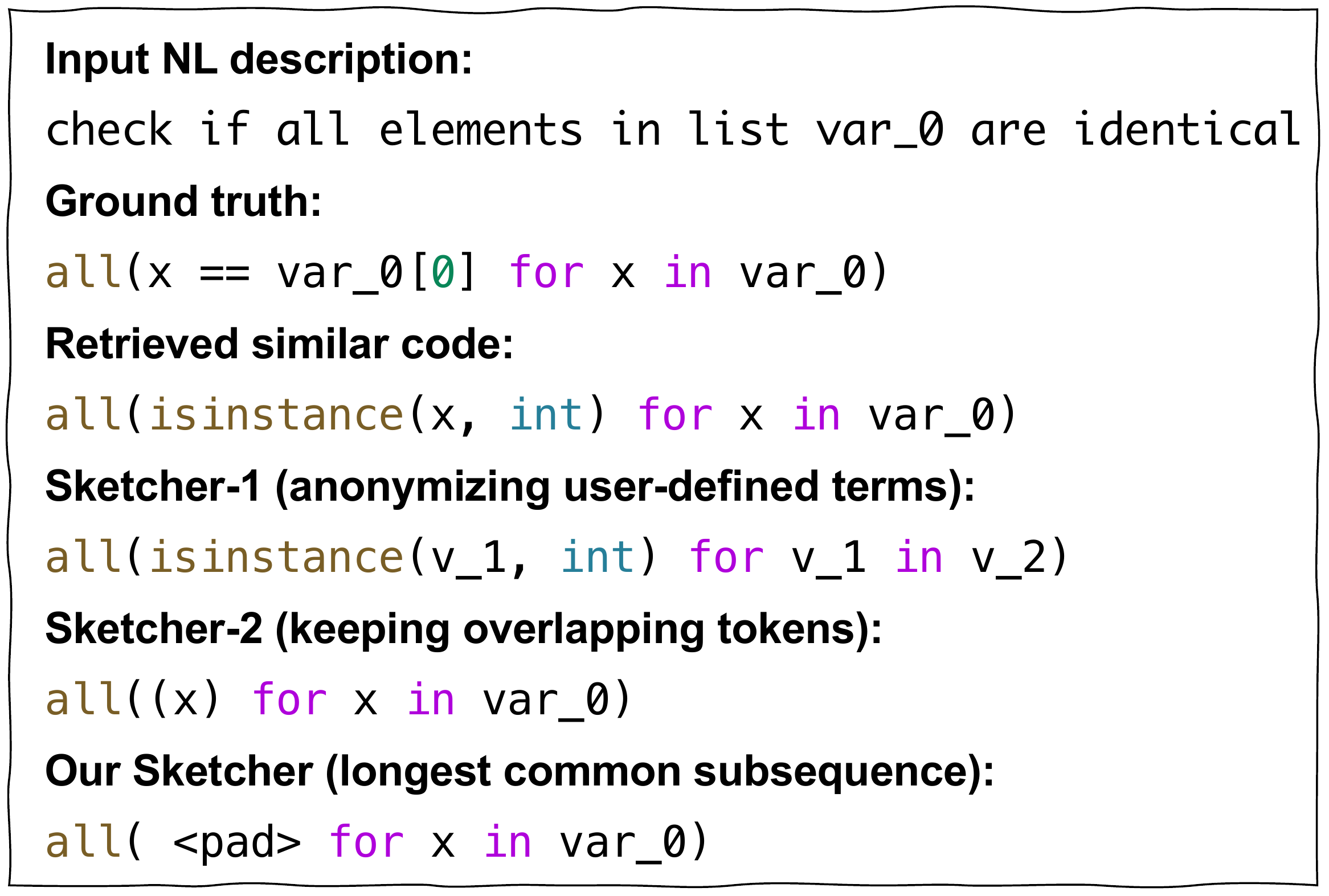}
\vspace{-0.3cm}
\caption{Examples of three sketches.}
\label{fig:sketcher}
\vspace{-0.5cm}
\end{figure}

% \begin{tcolorbox}[size=title]
% \textbf{RQ3: What is the better design choice for the code sketch?}
% \end{tcolorbox}

\noindent \textbf{RQ3: What is the better design choice for the code sketch?}

\textbf{Setup.} In this RQ, we provide three sketchers (\ie sketcher-1, sketcher-2, and our sketcher). The sketcher-1 utilizes a parse to anonymize the user-defined terms in the similar code (\ie string, constant, variable) and obtains a code sketch. The sketcher-2 trains a neural network to predict overlapping tokens between the similar code and the ground truth. The overlapping tokens are collected to build a sketch.
Our sketcher trains a neural network to predict the longest common subsequence (LCS) between the similar code and the ground-truth. The predicted LCS is viewed as a sketch. We present some examples of different sketchers in Figure \ref{fig:sketcher}.

\begin{figure*}[t]
\centering
\includegraphics[width=\linewidth]{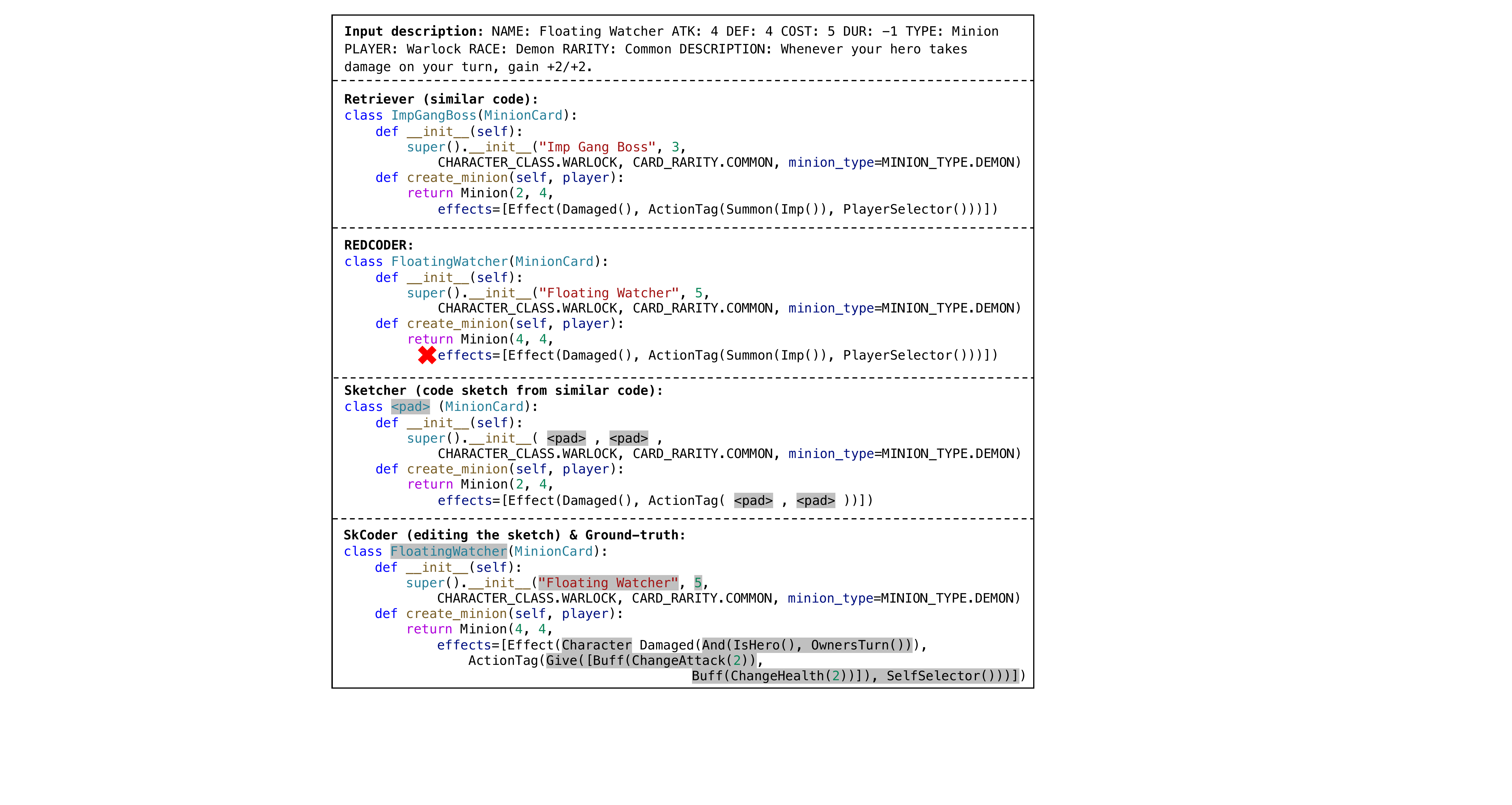}
\vspace{-0.5cm}
\caption{Examples of code snippets generated by different models. We highlight the parts that \method modifies on the sketch.}
\label{fig:case_study}
\vspace{-0.5cm}
\end{figure*}

\textbf{Results.} The experimental results are shown in Table \ref{tab:sketch}. We present the results of our \method with different sketchers and the result without a sketcher.

\textbf{Analyses.} 
\uline{(1) Introducing a sketcher can better utilize the retrieved code.} 
Compared to the model without a sketcher, the models with sketchers perform better. It
shows that the sketcher can better mine the knowledge from the retrieved code and our sketch-based approach is more promising than copy-based approaches.
\uline{(2) Our sketcher performs best among all baselines.} 
On both datasets, our sketcher brings 2x improvements (\eg 9.13\% vs. 0.43\%) over other sketchers.
This is because our sketcher can accurately extract the relevant content and leave irrelevant details, while other sketchers cannot.
As shown in Figure \ref{fig:sketcher}, sketch-1 outputs a sketch by anonymizing the user-defined terms. But the anonymized code still contains irrelevant parts (\eg \texttt{isinstance}) and even loses some reusable tokens (\eg \texttt{var\_0}). 
Sketch-2 only keeps tokens that may occur in the ground truth. It ignores the sequentiality of tokens, and the generated sketch probably is disorder and confusing (\eg \texttt{(x)}).
By contrast, the sketch produced by our sketcher is well-formed and provides a clear code structure.

% The placeholder (\ie \texttt{<pad>}) also specifies which parts may be further elaborated.

\begin{tcolorbox}[size=title]
\textbf{Answer to RQ3:} Code sketches are beneficial to reuse the knowledge in the retrieved code. Among multiple plausible sketchers, our sketcher performs best and brings a maximum of 9.13\% improvement in the EM.
\end{tcolorbox}

%% file: chapter/human_evaluation.tex
\vspace{-0.2cm}
\section{Human Evaluation}
\label{sec:human_evaluation}

The ultimate goal of code generation models is to assist developers in writing the source code. Thus, in this section, we conduct a human evaluation to assess our {\sc SkCoder}.

\textbf{Setup.}
Following previous work \cite{hao2022aixbench}, we manually evaluate the generated code by different models in three aspects, including \textit{correctness} (whether the code satisfies the given requirement), \textit{code quality} (whether the code does not contain bad code smell) and \textit{maintainability} (whether the implementation is standardized and has good readability). For each aspect, the score is integers, ranging from 0 to 2 (from bad to good). 
We randomly select 50 test samples and collect programs generated by 10 models on these samples. Finally, we obtain 500 programs (50*10) for evaluation. The evaluators are computer science Ph.D. students and are not co-authors. They all have programming experience ranging from 3+ years.
The 500 code snippets are divided into 5 groups, with each questionnaire containing one group. We randomly list the code and the corresponding input description on the questionnaire. Each group is evaluated anonymously by two evaluators, and the final score is the average of two evaluators' scores. Evaluators are allowed to search the Internet for unfamiliar concepts.

\begin{table}[t]
\caption{The results of human evaluation. All the p-values are substantially smaller than 0.005.}
\resizebox{\linewidth}{!}{
\begin{tabular}{lccc}
\toprule
Approach        & Correctness & Code quality & Maintainability \\ \midrule
GraphCodeBERT   & 0.9277      & 0.9872       & 1.3049          \\
CodeGPT         & 0.9798      & 1.0229       & 1.3306          \\
REDCODER *        & 1.0177      & 1.2038       & 1.5796          \\
CodeGen         & 1.1250      & 1.3610       & 1.5573          \\
PyCodeGPT       & 1.1098      & 1.3661       & 1.5442          \\
CodeParrot      & 0.9704      & 1.0814       & 1.3668          \\
GPT-Code-Clippy & 0.9646      & 1.0585       & 1.3672          \\
CERT-PyCodeGPT  & 0.9629      & 1.0439       & 1.3882          \\
CodeT5-base     & 1.1719      & 1.3908       & 1.5848          \\
\method         & \textbf{1.3705 ($\uparrow$ 16.95\%)}      & \textbf{1.5639 ($\uparrow$ 12.45\%)}      & \textbf{1.7764 ($\uparrow$ 12.09\%)}         \\ \bottomrule
\end{tabular}}
\label{tab:human_evaluation}
\vspace{-0.3cm}
\end{table}

\textbf{Results and Analyses.}
The results of the human evaluation are shown in Table \ref{tab:human_evaluation}. Our \method is better than all baselines in three aspects. Specifically, \method outperforms the SOTA model - CodeT5-base by 16.95\% in correctness, 12.45\% in code quality, and 12.09\% in maintainability. 
All the p-values are substantially smaller than 0.005, which shows the improvements are statistically significant.
The improvements prove the superiority of our \method in assisting developers in coding. Besides, we notice that the copy-based model - REDCODER performs well in maintainability but is poor in correctness and code quality. This is because REDCODER can generate natural programs by copying from the retrieved code. But some copied content is irrelevant and leads to incorrect code.

%% file: chapter/discussion.tex
% \vspace{-0.2cm}
\section{Discussion}
\label{sec:discussion}

\subsection{Case Study}
\label{sec:discussion:case_study}

Figure \ref{fig:case_study} presents some code snippets generated by different models on the HearthStone dataset. From the examples, we obtain the following findings. (1) The retrieved similar code provides a well-formed code structure and contains some irrelevant details (\eg \texttt{ImpGangBoss}). (2) As a copy-based approach, REDCODER wrongly repeat the inappropriate statement (\ie \texttt{effects=[Effect(...), ActionTag(...)])} without modifications. It causes the generated code is inconsistent with the input description.
(3) Our sketcher accurately keeps the relevant content and replaces irrelevant details with placeholders. The extracted sketch provides a clear start-point for editing. 
(4) Based on the input description, our \method further edits the sketch into the desired code. For example, the input description specifies the card's effect (\ie \texttt{whenever you hero takes damage on your turn, gain +2/+2}). \method modifies the \texttt{Damaged()} and \texttt{ActionTag()} calls in the sketch and adds more details (\eg \texttt{And(IsHero(), OwersTurn()}). Besides, the editor adds some components that are not in the sketch, such as \texttt{Character}.

\subsection{Threats to Validity}
\label{sec:discussion_threat}

There are three main threats to the validity of our work.

\textbf{The generalizability of our experimental results.}
To mitigate this threat, we carefully design the experimental datasets, metrics, and baselines. For the datasets, we follow previous studies \cite{yin2018tranx,hayati2018redcode,sun2020treegen} and pick three representative code generation datasets. 
The three datasets are collected from real software projects and communities and cover two popular programming languages (\ie Java and Python).
For the metrics, we select five widely used metrics, including the EM, BLEU, CodeBLEU, Pass@1, and AvgPassRatio. Existing work \cite{evtikhiev2022CGmetric} has proven the reliability of these metrics. 
To verify the superiority of our approach, we select 20 code generation models as our baselines for the comparison. They cover the most of representative work in the past six years. Besides, we run each approach three times and report the average results.

\textbf{The implementation of models.} 
It is widely known that deep learning models are sensitive to the implementation details, including hyper-parameters and network architectures. In this work, we need to implement baselines and our approach.
For the baselines, we use the source code provided by their original papers and ensure that the model's performance is comparable with their reported results.
For our approach, we implement a version that employs mainstream neural networks (details in Section \ref{sec:study_design:implementation}).
Due to the high training cost, we do a small-range grid search on several hyper-parameters (\ie learning rate and batch size), leaving other hyper-parameters the same as those in previous studies \cite{li2021editsum,guo2020graphcodebert,wang2021codet5}.
Thus, there may be room to tune more hyper-parameters and network architectures of our approach for more improvements.

\textbf{The impact of retrieved code.}
The retrieved code is an important element in our approach. Intuitively, when the retrieved code is less similar to the target code, the performance of our model may suffer. To address this threat, we have two thoughts. (1) A large-scale study on 13.2 million real code files found the proportion of reused code is up to 80\% \cite{mockus2007codereuse}. Therefore, we believe that it is quite possible to retrieve the similar code in real development scenarios.
(2) Even if the retrieved code is dissimilar to the target code, our \method can selectively focus on the retrieved code based on current requirements.
To prove this point, we randomly select code snippets from the retrieval corpus as the retrieved code and train a variant named {\sc SkCoder}-random.
The results are shown in Table \ref{tab:random_example}. {\sc SkCoder}-random has a drop compared to \method but still substantially outperforms CodeT5-base. It proves that our \method can adaptively extract valuable content from the retrieved code and has strong robustness.

\begin{table}[t]
\caption{The performance of {\sc SkCoder}-random.}
\centering
\begin{tabular}{lccc}
\toprule
Approach       & EM    & BLEU  & CodeBLEU \\
\midrule
CodeT5-base    & 28.91 & 80.46 & 73.11    \\
{\sc SkCoder}-random & 33.48 ($\uparrow$ 15.81\%) & 82.07 & 79.08    \\
\method        & 35.39 ($\uparrow$ 22.41\%) & 85.39 & 80.62   \\
\bottomrule
\end{tabular}
\label{tab:random_example}
\vspace{-0.4cm}
\end{table}

%% file: chapter/related_work.tex
\section{Related work}
\label{sec:related_work}

% 代码生成

% 代码生成就是从自然语言描述生成通用编程语言的任务。
Code generation aims to generate the source code that satisfies a given natural language description or requirement.
% 现有的工作可以分为：1）基于序列的方法；2）基于树的方法；3）基于预训练的方法
Existing work can be divided into three categories: sequence-based models, tree-based models, and pre-trained models.

% sequence-based models
\textbf{Sequence-based Models.} Sequence-based models treat the source code as a sequence of tokens and use neural networks to generate the source code token-by-token based on the input description.
Ling et al. \cite{ling2016latent} generate the source code with a structured attention mechanism to process the semi-structural input.
Hashimoto et al. \cite{hashimoto2018reedit} train a task-dependent retriever to retrieve the similar code, and then use the similar code as an additional input to the generator.
Wei et al. \cite{wei2019code} propose a code generation model based on dual learning, which performs better with the help of code summarization.

\textbf{Tree-based Models}
Program is strictly structured, and can be parsed into a tree, \eg Abstract Syntax Tree (AST).
Tree-based models generate a parse tree of the program based on the NL description and then convert the parse tree into the corresponding code.
Dong et al. \cite{dong2016seq2tree} generate the AST by expanding every non-terminal with an LSTM model.
Rabinovich et al. \cite{rabinovich2017asn} generate the AST with a decoder that has a dynamically-determined modular structure paralleling the structure of the output AST.
Yin et al. \cite{yin2018tranx} generate the tree-construction action sequence with an LSTM model, and construct the AST from the action sequence.
Sun et al. \cite{sun2020treegen} encode the natural language and grammar rules that have been generated with specially designed Transformer blocks, and predict the next grammar rule accordingly.

% pre-trained models
\textbf{Pre-trained Models}
Recent years have witnessed the emergence of pre-trained models \cite{CodeEditor, AceCoder, TiP}.
These models are pre-trained on massive data of source code and then fine-tuned on code generation task.
Pre-trained models can be divided into three categories.

(1) \textit{Encoder-only pre-trained models} only contains an encoder and is mainly used in code representation.
They are usually pre-trained with language comprehension tasks, \eg masked language modeling or replaced token detection. The recently proposed encoder-only pre-trained models include the CodeBERT \cite{feng2020codebert}, GraphCodeBERT \cite{guo2020graphcodebert}, etc.
(2) \textit{Decoder-only pre-trained models} are pre-trained to predict the next token based on the input context. GPT series \cite{radford2018gpt} are excellent decoder-only models for natural language processing, and there are many efforts to adapt similar ideas to code.
Lu et al. \cite{lu2021codexglue} adapt GPT-2 \cite{radford2019gpt-2} model on the source code, resulting in CodeGPT. Chen et al. \cite{chen2021codex} fine-tune GPT-3 \cite{brown2020gpt3} models on the code to produce CodeX and GitHub Copilot \cite{web:copliot}.
Neither CodeX nor GitHub Copilot is open-sourced, which leads to several attempts to replicate CodeX in industry and academia, resulting in CodeParrot \cite{web:codeparrot}, GPT-CC \cite{web:gpt-cc}, PyCodeGPT \cite{zan2022cert}, and CodeGen \cite{nijkamp2022codegen}.
CodeParrot and CodeGen are trained from scratch. PyCodeGPT and GPT-CC are fine-tuned from GPT-Neo\cite{black10gpt-neo}.
Zan et al. \cite{zan2022cert} propose a variant of PyCodeGPT. They first generate a sketch that anonymizes user-defined constants, and then generate the complete program from the NL and the sketch.
(3) \textit{Encoder-decoder pre-trained models} are composed of an encoder and a decoder. They can support both code representation and code generation tasks. 
Various successful encoder-decoder architecture in natural language processing has been transferred into the source code, resulting in powerful models, \eg CodeT5 \cite{wang2021codet5} and PLBART \cite{ahmad2021plbart}.

Inspired the code reuse, some studies introduce the similar code to augment code generation models.
Hayati et al. \cite{hayati2018redcode} retrieve the similar code with the input, and copy $n$-gram actions from the similar code during decoding. 
Hashimoto et al. \cite{hashimoto2018reedit} and Parvez et al. \cite{parvez2021redcoder} retrieve similar code snippets and feed them along with the input description to a generator.
They train the generator to learn to copy some reusable content from the similar code. We refer to these studies as copy-oriented approaches.
Different from copy-oriented approaches, our sketch-oriented \method mimics the developers' code reuse behavior, extracts content that is relevant to input requirement and ignores irrelevant parts in the similar code.
The extracted content is viewed as a code sketch and further edited to the target code with guidance of input requirement.

%% file: chapter/conclusion.tex
\section{Conclusion and Future Work}
\label{sec:conclusion}

During software development, human developers often reuse similar code snippets.
In this paper, we propose a novel sketch-based code generation approach named \method to mimic developers' code reuse behavior. 
Different from previous copy-based approaches, \method can extract the relevant content from the retrieved similar code and builds a code sketch.
The sketch is further edited by adding more requirement-specific details.
We conduct experiments on two public code generation datasets and a new Java dataset collected by this work. The new dataset contains 200k NL-code pairs and each test sample is equipped with a set of unit tests.
Experimental results show that \method substantially outperforms state-of-the-art baselines. The ablation study proves the effectiveness of code sketches and our approach is effective to different neural networks.
In the future, we will explore more effective sketchers and apply our sketch-based idea to large-scale pre-trained models. 